\documentclass[useAMS,usenatbib,usegraphicx]{mn2e}
\usepackage{lscape}
\title[Cyanopolyynes in hot cores: modelling G305.2+0.2]{Cyanopolyynes in hot cores: modelling G305.2+0.2}
\author[J.F. Chapman, T.J. Millar, M. Wardle,   M.G. Burton and  A.J. Walsh]
{J. F. Chapman$^{1}$
\thanks{E.mail: jchapman@physics.mq.edu.au},
T.J. Millar$^{2}$, M. Wardle$^{1}$ , M.G. Burton$^{3}$, A.J. Walsh$^{4}$ \\
$^{1}$ Department of Physics, Macquarie University, Sydney, NSW, 2109, Australia\\
$^{2}$ Astrophysics Research Centre, School of Mathematics and
Physics, Queen's University Belfast, Belfast BT7 1NN, UK\\
$^{3}$ School of Physics, University of New South Wales, Sydney, NSW, 2052, Australia\\
$^{4}$ School of Maths, Physics and IT, James Cook University, NSW,
4811, Australia}
\begin{document}

\date{Accepted 1988 December 15. Received 1988 December 14; in original form 1988 October 11}

\pagerange{\pageref{firstpage}--\pageref{lastpage}} \pubyear{2002}

\maketitle

\label{firstpage}

\begin{abstract}
We present results from a time dependent gas phase chemical model of
a hot core based on the physical conditions of G305.2+0.2. While the
cyanopolyyne HC$_3$N has been observed in hot cores, the longer
chained species, HC$_5$N, HC$_7$N, and HC$_9$N have not been
considered typical hot core species. We present results which show
that these species can be formed under hot core conditions. We
discuss the important chemical reactions in this process and, in
particular, show that their abundances are linked to the parent
species acetylene which is evaporated from icy grain mantles. The
cyanopolyynes show promise as `chemical clocks' which may aid future
observations in determining the age of hot core sources. The
abundance of the larger cyanopolyynes increase and decrease over
relatively short time scales, $\sim$10$^{2.5}$ years.  We present
results from a non-LTE statistical equilibrium excitation model as a
series of density, temperature and column density dependent contour
plots which show both the line intensities and several line ratios.
These aid in the interpretation  of spectral line data, even when
there is limited line information available.
\end{abstract}


\section{Introduction}
Hot cores are observed in regions of high mass star formation, and
are small, hot ($\sim 200\mathrm{K}$) and dense ($\sim
10^6\mathrm{cm}^{-3}$) clumps of gas (e.g., \citealt{S78, Pea83},
for a review see, e.g., \citealt{vdT03}). The chemistry of these
regions is unique, characterized by unusually high abundances of
saturated molecules, such as H$_2$O and NH$_3$ (e.g.,
\citealt{Hea87,Wea87,O93}). This unique chemistry is believed to be
a direct result of grain mantle evaporation due to an event, such as
the switching on of a nearby star or a source within the core itself
(e.g., \citealt{Kea98,WW96}). During an earlier high density phase
of collapse, these molecules are  formed on the surface of dust
grains (e.g., \citealt{WS91}),  released into the gas phase during
the onset of heating, and are processed into daughter species. Since
cores are both hot and dense, the time dependent gas phase chemistry
rapidly changes over time scales $<$100,000 years.  Species with
time dependent chemistry can be used as `chemical clocks' and may
help in determining the age of hot core regions (e.g.,
\citealt{MMG97}).

The origins of some interstellar molecules found in hot cores can be
separated into the following three groups. Those that are formed in
the cold gas:
\begin{eqnarray*}
\noalign{\hbox{e.g., CO, C$_2$H$_2$, C$_2$H$_4$, N$_2$, O$_2$, CS,}}
\end{eqnarray*}
 those formed on grain surfaces:
\begin{eqnarray*}
\noalign{\hbox{e.g., H$_2$O, CH$_4$, NH$_3$, H$_2$S,  CH$_3$OH,
H$_2$CO, C$_2$H$_5$OH, }}
 \noalign{\hbox{CO$_2$, HCOOH, NH$_2$CHO,
C$_2$H$_4$, C$_2$H$_6$,H$_2$CS, OCS,}}  \label{grainlist}
\end{eqnarray*}
and those formed in the hot gas (e.g., \citealt{Charn95})
\begin{eqnarray*}
\noalign{\hbox{e.g., CH$_3$OCH$_3$, CH$_2$CO, CH$_3$CHO,
HCOOC$_2$H$_5$,
 }}
 \noalign{\hbox{CH$_3$COCH$_3$, (C$_2$H$_5)_2$O,
CH$_3$OC$_2$H$_5$,  C$_2$H$_5$CN,  CH$_3$CN,}} \noalign{\hbox{HCN,
SO, SO$_2$, C$_2$H$_3$CN, HC$_3$N, CH$_3$NH$_2$, CH$_2$NH, SiO. }}
\label{gasphaselist}
\end{eqnarray*}

Observations of hot cores include molecular line  and dust continuum
emission, and some of the chemical core models, along with radiative
transfer calculations, have proven useful in reproducing observed
column densities (e.g., \citealt{MMG97,Kea98,Tea99,Dea02}). Chemical
models have been developed that include a pre-stellar phase or
initial conditions assuming the presence of evaporated grain mantle
species  (e.g., \citealt{BCM88,MHC91,CBM92, CHH93, MMG97,VW99,
Dea02,NM04,Wea04}). The injection of ice mantle material into the
gas phase is usually assumed to occur instantaneously since it is
driven by stellar radiation and the grain temperature rapidly
increases from 10K (in the cold phase) to $\sim200\mathrm{K}$ (in
the hot core) when the ices evaporate quickly. Gas-grain chemical
hot core models which assume simple molecules like acetylene,
C$_2$H$_2$, are released from grain mantles,  have been used to
account for the observed abundances of many O and N bearing
molecules (e.g., \citealt{CBM92}).

Carbon chain molecules make up a significant fraction of the
observed interstellar molecules and cyanopolyynes HC$_n$N are often
detected. The longer chain cyanopolyynes HC$_{2n+1}$N (n=2-5) are
associated with regions such as dark dust clouds, and are not
usually  associated with hot cores. HC$_{11}$N has been detected in
the interstellar medium \citep{Betal97}, with the shorter chained
species, for example HC$_3$N and HC$_5$N,  more commonly associated
with a number of sources  including circumstellar shells  and star
forming regions. Cyanodiacetylene, HC$_5$N, is believed to be formed
inefficiently in hot cores  since grain mantles do not comprise of
larger unsaturated molecules such as diacetylene, HCCCCH
\citep{Millar97}. Recently, \citet{Sakaietal08} detected the longer
chain cyanopolyynes HC$_5$N, HC$_7$N and HC$_9$N towards the
low-mass protostar IRAS 04368+2557 L1527. Using a gas grain model,
\citet{HHG08} modelled the chemistry of L1527, and determined that
the composition of the cloud is most likely a result of warm, 30K,
chemistry than cold processes.

The rotational lines from vibrational excited states of some
cyanopolyynes have been detected in hot core sources.
\citet{Wyroetal99} observed a sample of six hot core sources,
including G10.47+003, in vibrationally excited lines of HC$_3$N.
Observations of the Orion Molecular cloud Core (OMC-1) have included
the J $= 12 \rightarrow$ 11 transition of the v$_6$ = 1
vibrationally excited state \citep{Goldetal83, Goldetal85}, and  5
transitions of the v$_7$ = 1 vibrationally excited state  of
HC$_3$N. The higher cyanopolyynes are not usually detected, however,
10 rotational transitions of HC$_5$N  and 8 transitions of HC$_7$N
\citep{Turner91} have been observed in the Orion core (OMC-1).
HC$_5$N (\citealt{Averyetal79, Turner91}) and HC$_7$N
\citep{Turner91} have also been detected in the molecular cloud of
Sgr B2.




Models of the gas phase chemistry are extremely useful as predictive
tools with regards to observations. However, extracting molecular
abundances from line emission, suitable for comparison with the
model predictions is not a straight forward task. Furthermore,
although theoretical models can produce abundances for hundreds of
molecules, observations are limited to specific frequency bands, and
possibly contain line emission from only a handful of species. A
broad understanding of the chemistry is thus not often possible with
limited molecular line data for a specific source.



Modelling tools to aid in the interpretation  of spectral line data
are especially important for large line surveys, and efficient
methods are needed to obtain  physical and chemical parameters, such
as column densities, relatively easily even with large numbers of
species and lines. Currently, collisional rate coefficients are
available for approximately 24 species  on the Leiden atomic and
molecular database (LAMDA) (\citealt{Setal05}), with several others
in  literature (e.g., CH$_3$CN in \citealt{green86}), and so
statistical equilibrium calculations are limited to these. Moreover,
observations are limited in frequency, and so limit the process
further. The list of species for which collisional rate coefficients
are available include CS, OCS, SO, SO$_2$, SiO, HCO$^+$, HC$_3$N,
HCN, HNC, H$_2$O, OH, CH$_3$OH, and NH$_3$. There are many more
species of astronomical interest that do not appear in the LAMDA
database, or elsewhere in the literature.

The aim of this paper is to model the gas phase chemistry of the hot
core associated with G305.2+0.2 with particular focus on the
formation of the cyanopolyynes. Non-LTE calculations are used to
examine the excitation properties of these species. Section
\ref{chemicalmodelsection} outlines the chemical model with the
cyanopolyyne HC$_{(3,5,7,9)}$N  chemistry discussed in Section
\ref{chemicalmodelsection-cyan}. We reveal that the longer chain
cyanopolyynes can  be produced in appreciable abundances assuming
the physical conditions of G305.2+0.2. In Section
\ref{statequilsection} we present results from a statistical
equilibrium model which produces  integrated  line intensities. We
present a series of density $\mathrm{n_{H2}}$, temperature T and
column density N$_j$ dependent contour plots which will directly aid
in the interpretation of observations. Non-detections of HC$_5$N and
HC$_7$N from \cite{Walshetal07} in G305.2+0.2 are explained using
the non-LTE model.

\begin{table}
 \centering
 \begin{minipage}{90mm}
  \caption{The initial gas phase abundances, with respect to H nuclei, assumed in the chemical model. }
 \label{tableinitialabundance}
  \begin{tabular}{llll}
  \hline
Species   & Abundance & Species   & Abundance   \\
\hline \\
CH$_3$OH & $5\times10^{-7}$ & C$_2$H$_2$  & $5\times 10^{-7}$   \\
NH$_3$ &  $1 \times 10^{-5}$ & C$_2$H$_4$ & $5 \times 10^{-9}$    \\
H$_2$S & $1 \times 10 ^{-6}$& C$_2$H$_6$  & $5 \times 10^{-9}$    \\
H$_2$O & $1 \times 10 ^{-5}$& CH$_4$      & $2 \times 10^{-7}$   \\
O$_2$  & $1 \times 10^{-6}$ & CO$_2$      & $5.0 \times 10^{-6}$  \\
H$_2$CO & $4 \times 10 ^{-8}$& OCS        & $5.0 \times 10^{-8}$ \\
CO     & $5 \times 10^{-5} $&  N$_2$      & $2 \times 10^{-7}$  \\
He$^+$     & $2.5 \times 10^{-11}$ & H$^+$& $1 \times 10^{-10} $ \\
H$_3^+$    &$ 1 \times 10^{-8}$ &  Si & $3.6 \times 10^{-8}$ \\
Fe$^+$       & $ 2.4 \times 10^{-8}$&\\
 \hline
\end{tabular}
\end{minipage}
\end{table}

\section{Chemical Model}\label{chemicalmodelsection}



\begin{table*}
 \centering
  \caption{Species present in model calculations, grouped into the  number  n of atoms.}\label{specieslist}
    \begin{tabular}{lllllllll}
   \hline

\bf{n = 1} &  H  &  He  & N & O   & C  & S  & Si  & Fe   \\
           &  H+ &  He+ & N & O+  & C+ & S+ & Si+ & Fe   \\
   \hline

\bf{n = 2} & HS     & CH  &  NH  & N2  & NO  &  NS  & CN  & C2 \\
           & HS+    & CH+ &  NH+ & N2+ & NO+ &  NS+ & CN+ & C2+     \\
           & CO     & CS  &  O2  & OH  & SO  &  S2  & SiS & H$_2$+\\
       & CO+    & CS+ &  O2  & OH+ & SO+ &  S2+ & SiS+ &   \\
 \hline
\bf{n = 3} &   H$_2$O & CH$_2$    &  NH$_2$ &  HS$_2$   & C$_2$H  & H$_2$S+  &  CO$_2$ &  C$_2$S    \\
           &   H$_2$O+& CH$_2$    &  NH$_2$+&  HS$_2$+  & C$_2$H+ & H$_2$S   &  CO$_2$+&  C$_2$S+  \\
           &   C$_2$O  &  SO$_2$  &  C$_3$  &  HCO      &  HCS    & HSO+    & HNS+     &  HSiS+     \\
           &   C$_2$O+ &  SO$_2$+ &  C$_3$+ &  HCO+     &  HCS+   &C$_2$N+  & HN$_2$+  &  H$_3$+  \\
           &    HCN    & OCS      &  HNC    &  HSO+ & HSiS+  \\
           &    HCN+   & OCS+     &  CNC+   &  HNS+     \\
\hline
\bf{n = 4} &   NH$_3$  &  CH$_3$  & C$_2$H$_2$   &  H$_2$CO  &  C$_3$H  &  H$_2$CS  & C$_4$  & C$_3$O+  \\
       &   NH$_3$+ &  CH$_3$+ & C$_2$H$_2$+  &  H$_2$CO+ &  C$_3$H+ &  H$_2$CS+ & C$_4$+ &  C$_3$O  \\
           &   C$_3$N  &  C$_3$S   &  H$_2$S$_2$  &  HOCS+   &  H$_2$NC+ & HCNH+    & HC$_2$O+ &HSO$_2$+ \\
       &   C$_3$N+ &  C$_3$S+  &  H$_2$S$_2$+ &  HC$_2$S+ &  H$_3$O+  & H$_3$S+  &  HCO$_2$+       \\
\hline
\bf{n = 5} & HC$_3$N    & CH$_4$  &  C$_4$H  & C$_2$H$_3$  &  C$_5$   & C$_4$S  &  H$_2$CCC & C$_3$H$_2$  \\
       & HC$_3$N+   & CH$_4$+ &  C$_4$H+ & C$_2$H$_3$+ &  C$_5$+  & C$_4$S+ &  CH$_2$CO+ & C$_3$H$_2$+  \\
        & NH$_4$+  &  C$_4$N+  &  HCOOH   &   HC$_3$O+  &  H$_3$CS    & CH$_2$NH  & H$_2$CCO+& HC$_3$S+ \\
       & H$_3$CO+   & H$_3$S$_2$+ \\
\hline

\bf{n = 6} &C$_5$H & C$_2$H$_4$ & CH$_3$OH   &    C$_6$   &  C$_5$N & CH$_3$CN & C$_3$H$_3$   &  H$_2$CCCC    \\
     &  C$_5$H+    & C$_2$H$_4$+ &CH$_3$OH+  &  C$_6$+  &  C$_5$N+  & CH$_3$CN+& C$_3$H$_3$+  & C$_3$H$_2$O+ \\
     &  CH$_5$+ & HC$_4$N+ & CH$_2$NH$_2$+     & HC$_4$S+   &  H$_2$C$_3$H+ &     HC$_3$NH+   &  CH$_3$CO+ &HCOOH$_2$+ \\
     &  C$_4$H$_2$+ \\
 \hline
\bf{n = 7} &  C$_6$H  &  HC$_5$N   &  CH$_3$CCH    & C$_2$H$_5$  &  C$_7$  & CH$_3$CHO  &C$_5$H$_2$   &   CH$_3$OH$_2$+  \\
           &  C$_6$H+ &    HC$_5$N+   &  CH$_3$CCH+& C$_2$H$_5$+   &C$_7$+ & CH$_3$CHO+ & C$_5$H$_2$+   &  CH$_3$CNH+ \\
       &   C$_4$H$_3$+ & H$_2$C$_4$N+ &  CH$_2$CHCN&  H$_3$C$_3$O+   &    \\
\hline
\bf{n = 8} &  C$_7$H  &  C$_8$   & CH$_3$C$_3$N & C$_6$H$_2$  &  C$_7$N &    CH$_3$CH$_3$  & HCOOCH$_3$  & C$_3$H$_5$+  \\
       &  C$_7$H+ &  C$_8$+  &  CH$_3$C$_3$N+& C$_6$H$_2$+   &  C$_7$N+ &      CH$_3$CH$_3$+&   COOCH$_4$+ & H$_2$C$_5$N+    \\
           & C$_5$H$_3$+ & C$_6$H$_3$+   & CH$_3$CHOH+  &  \\
\hline
\bf{n = 9 } & C$_8$H  & HC$_7$N & CH$_3$OCH$_3$  & C$_7$H$_2$ &   C$_2$H$_7$+ &       CH$_3$C$_4$H  & C$_2$H$_5$OH   &  H$_5$C$_2$O$_2$+  \\
            & C$_8$H+ & HC$_7$N+ &  CH$_3$OCH$_3$+ & C$_7$H$_2$+& C$_4$H$_5$+ &     CH$_3$C$_4$H+ & C$_2$H$_5$OH+ &CH$_3$C$_3$NH+  \\
            &  C$_9$ \\
\hline

\bf{n $\ge$ 10} & C$_9$H &  C$_9$N & CH$_3$COCH$_3$   &  C$_8$H$_2$  &  HC$_9$N  &  C$_9$H$_2$ & CH$_3$C$_7$N  & C$_2$H$_5$OH$_2$+\\
                & C$_9$H+ & C$_9$N+ & CH$_3$COCH$_3$+&  C$_8$H$_2$+  &  HC$_9$N+ &  C$_9$H$_2$+ &CH$_3$OCH$_4$+  &  H$_2$C$_7$N+\\
               & C$_5$H$_5$+&   C$_7$H$_3$+  &  CH$_3$COCH$_4$+      &  C$_9$H$_3$+   & C$_8$H$_3$+& H$_2$C$_9$N+    \\
\hline
 Conserved species: &  e$^-$   &  H$_2$     &      &     \\
        \hline
  \end{tabular}
\end{table*}

The physical model for the core is  similar to that of \cite{MMG97}
and \cite{Tea99}  which assume a spherical core geometry. The
density and temperature is constant throughout since we want to
examine the general gas phase chemistries. We assume a density of
n$_\mathrm{H2}$ = 8 $\times$$10^4$cm$^{-3}$ and a temperature of
200K, which was derived for the hot core of G305.2+0.2, G305A, in
\cite{Walshetal07}, with a radius of 1.1pc at a distance of 3.9kpc.
The current modelling work uses the latest chemical rate
coefficients from \cite{Wetal05}. The model results of
\cite{Walshetal07} used the earlier \cite{TMM00} version. Also, the
current modelling work includes the longer chain cyanopolyynes,
HC$_7$N and  HC$_9$N, along with associated species needed for their
formation. These species were not considered previously in
\cite{Walshetal07}.

The initial gas phase abundances are given in Table
\ref{tableinitialabundance} (e.g., \citealt{MMG97,Walshetal07}). The
initial abundances are based on previous models of hot cores (e.g.,
\citealt{MMG97}, \citealt{Rodandcharn01},\citealt{Walshetal07}) and
the mean interstellar values. The source is assumed to fill the
beam, and so beam dilution does not play a large role in the model
results. The model contains 245 species, consisting of the 8
elements H, He, C, N, O, Si, S, and Fe. The reaction rate network
contains 3184 reactions.
 Included in the model are the cyanopolyynes HC$_{2n+1}$N (n=0-4),
the carbon chained molecules C$_n$H$_2$ (n=3-9) and the hydrocarbon
chains C$_n$H (n=1-9). The full set of species are given in Table
\ref{specieslist} with the initial mantle abundances in Table
\ref{tableinitialabundance}. The species list includes many
previously observed hot core species along with essential species
needed for their production.

Discrepancies between models of dense interstellar clouds
\citep{MLH87} highlight the sensitivity of astrochemical models  to
the reaction rate coefficients as well as the reaction network. The
reaction rates are taken from the most recent UMIST database which
includes the latest laboratory data \citep{Wetal05}. The reaction
networks of some of the key core species will be discussed in more
detail below. Many of the rate coefficients are temperature
dependent. There are a select few reactions  with rate coefficients
that are ill-defined outside their prescribed temperature range,
with rate coefficients exponentially increasing with decreasing
temperature. In these cases the rate coefficients are kept constant
at the lowest or highest temperature range value nearest the
temperature of interest.


Numerically, the core is described by a series of 22 concentric
shells, with a density and temperature assigned to each.  At each
depth point or shell, the initial abundances are set and the
reaction network is calculated over 10$^8$ years. However, the
chemistry rapidly evolves over much shorter time scales than this.
The species abundances are integrated along the line of sight and
the column density of a species $j$ at time $t$ is
\begin{eqnarray}
N_j(r,\theta,t) =\int_0^\infty n_j(s,t)ds
\end{eqnarray}
where $n_j(s,t)$ is the species abundance and $s$ is the depth
through the core along the line of sight. The coordinates $r$ and
$\theta$ specify the position of the pencil beam with respect to the
centre of the core. We approximate the telescope beam using a
Gaussian with an area normalized primary beam:
\begin{eqnarray}\label{gaussian}
P(r) = \frac{1}{2\pi\sigma^2} \exp{(-\frac{r^2}{2\sigma^2})},
 \,\,\,\,\,\sigma = \frac{\mathrm{HPBW}}{2\sqrt{2\mathrm{In}2}}
\end{eqnarray}
where $r$ is the radius measured from the centre of the gaussian and
HPBW is the half power beam width. $P(r)$ is normalized so that
\begin{eqnarray}
\int_{0}^{2\pi}\int_{0}^{\infty} P(r) r dr d\theta =1.
\end{eqnarray}

To evaluate the gaussian beam, column densities are firstly
evaluated along pencil beams on grid points through the chemical
core model. These column densities are then convolved with the
Gaussian, equation (\ref{gaussian}), giving  the beam-averaged or
beam diluted column density $N_j(t)$
\begin{eqnarray}
N_j(t)=\frac{1}{2\pi\sigma^2}\int_0^{2\pi}\int_0^\infty
N_j(r,\theta,t)\exp\left( -\frac{r^2}{2\sigma^2}\right)rdrd\theta.
\end{eqnarray}

\section{Cyanopolyynes}\label{chemicalmodelsection-cyan}

In a related project, we have begun a survey for HC$_5$N in
methanol-maser selected hot molecular cores from the
\citealt{Purcellea06} list, using the Tidbinbilla 34m telescope to
measure the 31.95 GHz J=12-11 line. These results will be reported
separately, when the survey is complete. However, we can report that
HC$_5$N is frequently observed in the cores selected from this list,
typically when CH$_3$CN 5-4 92 GHz line emission was also detected by
\citealt{Purcellea06}. Sources have been found to contain both
CH$_3$CN and HC$_5$N, or neither at all, suggesting their possible
use as chemical clocks. We will explore the chemical pathways below
and we show that HC$_5$N may well be a hot core species. We also
discuss the likely role of other carbon bearing molecules.

\cite{Fuketal98} have demonstrated the importance of these
neutral-neutral reactions:
\begin{equation}\label{cyanreac}
\mathrm{C_{2n}H_2+ CN\rightarrow HC_{2n+1}N+H},
\end{equation}
in the formation of cyanopolyynes in the ISM. Cyanoacetylene,
HC$_3$N is often detected in cold gas in molecular clouds, and is
formed efficiently via the reaction:
\begin{equation}\label{HC3nreac}
\mathrm{C_2H_2 } +\mathrm{CN} \rightarrow \mathrm{HC_3N}
+\mathrm{H},
\end{equation}
with a rate coefficient from \citealt{smithetal04}.  Observed
abundances of C$_2$H$_2$, (as well as CH$_3$CN and H$_2$CO) cannot
be produced in the hot core gas phase alone, but can be accounted
for in the cold phase chemistry where they are frozen out onto the
grains before being released into the gas phase (e.g.,
\citealt{BCM88,charnandtiel92,CBM92}). \cite{LahuisVandis00} found
high abundances of hot C$_2$H$_2$ towards massive YSOs with ISO and
concluded that the most likely origin is the evaporation of grain
mantles.  Maps of C$_2$H$_2$ emission at 13.7 microns with Spitzer
toward Cepheus A East in \cite{Sonnetal07} show that the
distribution is likely due to the sputtering of acetylene in icy
grain mantles. Although C$_2$H$_2$ has yet to be directly detected
in interstellar ices, the upper limit is high, 10$^{-5}$ with
respect to hydrogen \citep{boulinetal98}, and the fractional
abundance of C$_2$H$_2$ in Table \ref{tableinitialabundance} is
chosen to be below this limit. \cite{Rodandcharn01} have also
demonstrated with a gas grain model that high abundances of
C$_2$H$_2$ may be produced in the gas phase provided that a large
enough abundance of methane is ejected from the grain mantles.
Inside hot cores, C$_2$H$_2$  is released from the grain mantles and
reaction (\ref{HC3nreac}) proceeds  relatively easy for hot core
temperatures (100-300K).




\begin{figure}\label{figure1}
\includegraphics[width=85mm]{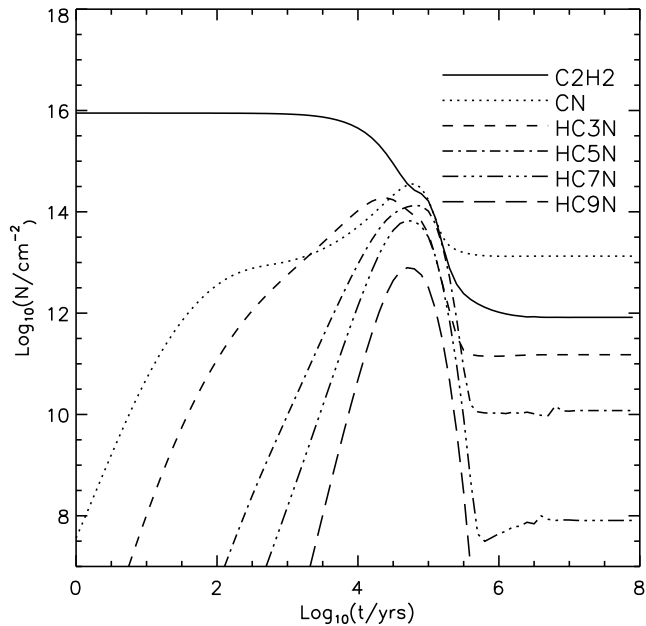}
 \caption{ Time dependent column densities for the chemical model of Section \ref{chemicalmodelsection-cyan} with n$_\mathrm{H2}$ = 8 $\times$$10^4$cm$^{-3}$ and a
temperature of 200K. The different length cyanopolyynes are
compared, along with the parent species CN and
C$_2$H$_2$.}\label{figure1}
\end{figure}

Results from the time dependent gas phase chemical model described
in Section \ref{chemicalmodelsection} are shown in Figure
\ref{figure1}. The column density of HC$_3$N rises to a peak above
N=$10^{14.5}$cm$^{-2}$ at t=10$^{4.3}$ years, and reaches its steady
state abundance after 10$^6$ years. The abundance of HC$_3$N
initially increases with CN since it is mainly produced via the
reaction of CN with C$_2$H$_2$, reaction
 (\ref{HC3nreac}). Also plotted are the  larger
cyanopolyynes HC$_{5,7,9}$N, whose abundances also peak near
t=10$^{4.3}$ years.  These species are not usually considered hot
core species but in Figure \ref{figure1} they are all reach
appreciable column densities over N = 10$^{12.5}$ cm$^{-2}$. The
peak column densities of the shorter molecules, HC$_5$N and HC$_7$N,
are one order of magnitude  larger than that of HC$_9$N.   Also, all
species reach steady state abundances after 10$^6$ years.

In the ISM cyanodiacetylene, HC$_5$N is mainly formed in the
neutral-neutral reaction
\begin{equation}\label{formHC5N}
\mathrm{CN} + \mathrm{H_2CCCC} \rightarrow \mathrm{HC_5N + H}.
\end{equation}
The rate coefficient for this reaction is $2.55 \times 10^{-10}$
cm$^3$ s$^{-1}$ at a temperature of 300 K \citep{Wetal05}. This was
confirmed as the dominant reaction in the  production of HC$_5$N in
the hot core results of  Figure \ref{figure1}.  This reaction has
not been considered important in the past since it was not thought
that H$_2$CCCC was  formed on the grain surfaces or very efficiently
in the gas phase. However, H$_2$CCCC is formed efficiently via
\begin{equation}\label{C4H2reac}
\mathrm{C_2H + C_2H_2 \rightarrow H_2CCCC} + \mathrm{H}
\end{equation}
with a rate coefficient of  $1.5\times 10^{-10}$ cm$^3$ s$^{-1}$ at 200K.
 The column densities of H$_2$CCCC and C$_2$H   are
plotted in Figure \ref{figure2}, reaching peak column densities of
10$^{14.8}$cm$^{-2}$ and  10$^{14.5}$cm$^{-2}$, respectively, near
10$^4$ years. Therefore, the abundance of H$_2$CCCC, and thus
HC$_5$N, also depend on acetylene C$_2$H$_2$. CN is produced in
relatively large quantities, furthering the efficiency of reaction
(\ref{formHC5N}). CN is produced mainly by dissociative
recombination of H$_2$NC$^+$,
\begin{eqnarray}
\mathrm{H_2NC^+}  + \mathrm{e^-} &\rightarrow&
\mathrm{CN}+\mathrm{H_2} ,
\end{eqnarray}
and at later times after 10$^5$ years by the reaction
\begin{eqnarray}\label{CN2reac}
\mathrm{C} +\mathrm{NO} &\rightarrow& \mathrm{CN} + \mathrm{O}.
\end{eqnarray}
H$_2$NC$^+$ is formed by
\begin{eqnarray}
\mathrm{He^+} +\mathrm{CO} &\rightarrow& \mathrm{O} +
\mathrm{C^+} +\mathrm{He}\\
\mathrm{C^+} +\mathrm{NH_3} &\rightarrow& \mathrm{H_2NC^+} +
\mathrm{H},
\end{eqnarray}
and reaction (\ref{CN2reac}) is preceded by
\begin{eqnarray}
\mathrm{N} +\mathrm{OH} &\rightarrow& \mathrm{NO} +
\mathrm{H}\\
\mathrm{H_3O^+} +\mathrm{e^-} &\rightarrow& \mathrm{OH} + \mathrm{H}
+\mathrm{H}.
\end{eqnarray}

Carbon chained polar molecules of the form H$_2$C=C..C have been
detected in the laboratory and have been identified  as useful
species to radio astronomers being  stable under interstellar
conditions. \cite{Killianetal90} have detected propadienlylidene
H$_2$CCC and butatrienylidene H$_2$CCCC in the laboratory, and
suggest that the H$_2$C$_n$ cumulene carbenes may be an another
important sequence of interstellar molecules, comparable with the
cyanopolyynes (HC$_n$N observed in space up to $n=11$) and the
linear hydrocarbon chains C$_n$H (observed up to n=8). H$_2$CCCC is
an isomer of diacetylene HCCCCH and has been detected in the dark
cloud TMC-1 \citep{Kawetal91} and the circumstellar shell of IRC
+10216 \citep{Cernetal91}.

The increase of the abundance of HC$_7$N  in Figure \ref{figure1} is
due to the reaction network
\begin{eqnarray}\label{HC7Nnetwork2}
\mathrm{C_2H_2} + \mathrm{C_4H} &\rightarrow& \mathrm{C_6H_2}
+\mathrm{H}  \label{HC7Nnetwork1} \\
\mathrm{CN} + \mathrm{C_6H_2} &\rightarrow& \mathrm{HC_7N + H}.
\label{HC7Nnetwork3}
\end{eqnarray}
Once again the abundance of the cyanopolyyne HC$_7$N is directly
related to acetylene. Similarly, the increase in the abundance of
HC$_9$N can be accounted for by the reactions
\begin{eqnarray}
\mathrm{C_6H_2} + \mathrm{C_2H} &\rightarrow& \mathrm{C_8H_2} +\mathrm{H}  \\
\mathrm{CN} + \mathrm{C_8H_2} &\rightarrow& \mathrm{HC_9N + H},
\end{eqnarray}
and this can be related to the HC$_7$N reaction network (reaction
\ref{HC7Nnetwork1}) via the C$_6$H$_2$ molecule. The column
densities of the  molecules responsible for the production of the
cyanopolyynes are plotted in Figure \ref{figure2} and are compared
with the daughter species HC$_5$N, HC$_7$N and HC$_9$N in Figure
\ref{figurecomparecyan-daughter}. The abundance of H$_2$CCCC
increases to a peak of 10$^{14.8}$cm$^{-2}$ at 10$^4$ years. The
abundance of the  daughter species HC$_5$N also increases reaching a
peak of 10$^{14.2}$cm$^{-2}$ near 10$^5$ years. The peak in HC$_5$N
occurs at a later time then its parent H$_2$CCCC.  Similar
relationships between C$_6$H$_2$ and HC$_7$N, and between C$_8$H$_2$
and HC$_9$N, can also be seen.

\begin{figure}\label{figurecomparecyan-daughter}
\includegraphics[width=85mm]{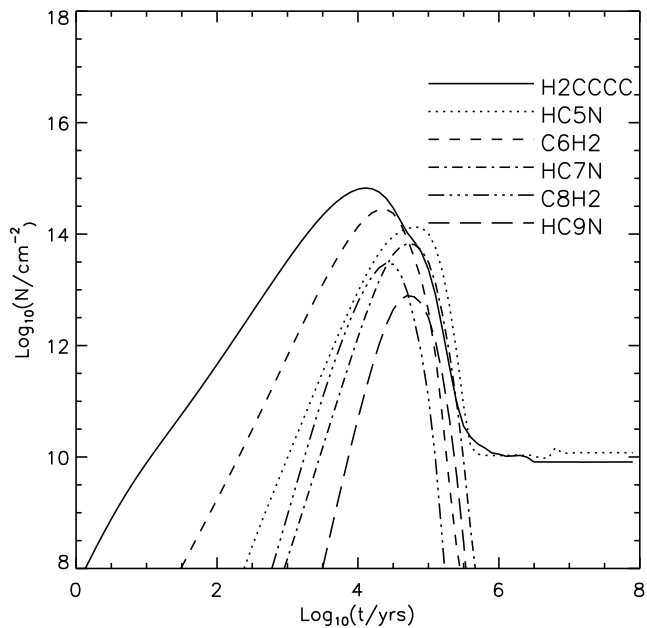}
\caption{As for Figure  \ref{figure1}, but for a comparison of the
higher order cyanopolyynes HC$_n$N (n$>$3) with their parent
species.}\label{figurecomparecyan-daughter}
\end{figure}

The formation of the cyanopolyynes, HC$_5$N , HC$_7$N and HC$_9$N,
are sensitive to the abundances of CN and the carbon chain molecules
H$_2$CCCC, C$_6$H$_2$, and C$_8$H$_2$. Furthermore, the abundance of
the carbon chain molecules are dependent on acetylene C$_2$H$_2$.
Thus one would expect, that if C$_2$H$_2$  is present in the gas
then the larger cyanopolyynes may also be present. Figure
\ref{figure3} illustrates this, showing the resulting abundances
when no C$_2$H$_2$ is initially evaporated from grain mantles. The
acetylene abundance initially increases to a column density $\sim
10^{14}$cm$^{-2}$ near 10$^{4.3}$ years. HC$_3$N, HC$_5$N, HC$_7$N
and HC$_9$N reach peak densities of $10^{13.5}$cm$^{-2}$,
$10^{12.7}$cm$^{-2}$, $10^{11.8}$cm$^{-2}$, and
$10^{10.4}$cm$^{-2}$, respectively. These column densities  are a
magnitude (or in some cases less) smaller then those of Figure
\ref{figure1}.

\begin{figure}\label{figure2}
\includegraphics[width=85mm]{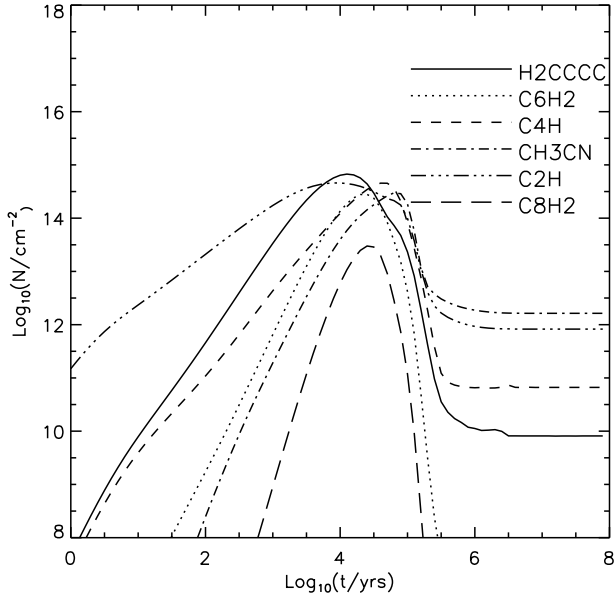}
 \caption{As for Figure  \ref{figure1}, but for the parent species of the
higher order cyanopolyynes HC$_n$N (n$>$3).}\label{figure2}
\end{figure}

\begin{figure}\label{figure3}
\includegraphics[width=85mm]{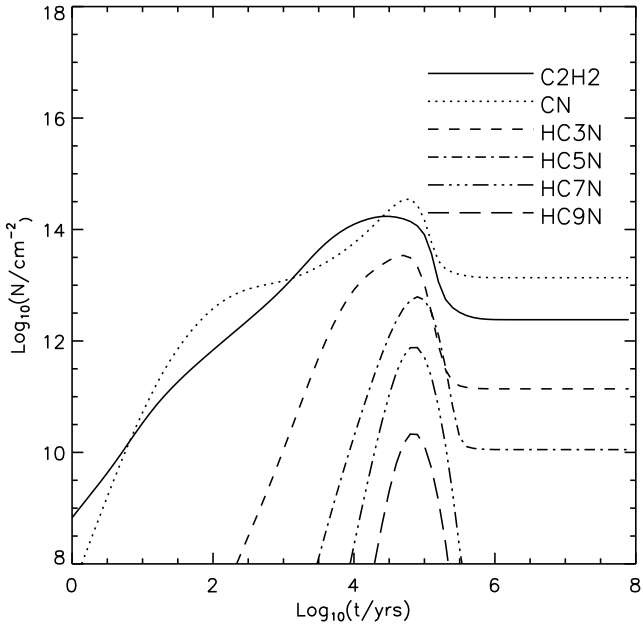}
 \caption{As for Figure  \ref{figure1}, however, C$_2$H$_2$ is
assumed to form in the gas only, with no initial abundance related
to grain mantle evaporation. }\label{figure3}
\end{figure}

Since CH$_3$CN emission is often used as a signature of the presence
of a hot molecular core, we have investigated its relative abundance
compared to HC$_5$N as a function of time.  CH$_3$CN is efficiently
produced via
\begin{eqnarray}
\mathrm{CH_3CNH^+}  + \mathrm{e^-} &\rightarrow& \mathrm{CH_3CN + H}
\label{CH3CNreac}
\end{eqnarray}
and the abundance of CH$_3$CNH$^+$ is driven by  radiative
association:
\begin{equation}
\mathrm{CH_3^+} + \mathrm{HCN} \rightarrow  \mathrm{CH_3CNH^+} +
h\nu. \label{CH3CNHreac}
\end{equation}

The methyl cyanide chemistry is shown in Figure \ref{figure4}. For
the first 10$^4$ years, CH$_3^+$ is produced by
\begin{eqnarray}
\mathrm{{H_3}^+} +  \mathrm{CH_3OH} &\rightarrow&   \mathrm{CH_3^+}
+ \mathrm{H_2O} + \mathrm{H_2}
\end{eqnarray}
and
\begin{eqnarray}
\mathrm{H_2} + \mathrm{CH_2^+} &\rightarrow& \mathrm{CH_3^+} + \mathrm{H}
\end{eqnarray}
thereafter.
The HCN abundance is driven by
\begin{eqnarray}
\mathrm{NH_3} +  \mathrm{HCNH^+} &\rightarrow& \mathrm{HCN} +
\mathrm{NH_4^+}\label{HCN1}
\end{eqnarray}
for the first $10^5$ years and
\begin{eqnarray}
 \mathrm{HCNH^+}   +  \mathrm{e^-}
&\rightarrow& \mathrm{HCN} + \mathrm{H} \label{HCN2}
\end{eqnarray}
after 10$^5$ years. These are preceded by
\begin{eqnarray}
\mathrm{H_2}   +  \mathrm{HCN^+}&\rightarrow& \mathrm{HCNH^+} + \mathrm{H} \\
\mathrm{C^+}   +  \mathrm{NH_3}&\rightarrow& \mathrm{HCN^+} + \mathrm{H_2} \\
\mathrm{C^+}   +  \mathrm{NH_2}&\rightarrow& \mathrm{HCN^+} + \mathrm{H}.
\end{eqnarray}

In Figure \ref{figure4}, the destruction of the mantle species
NH$_3$ and CH$_3$OH begins near 10$^4$ years, signalling the start
of the rapid gas phase chemistry between 10$^4$ and 10$^5$ years. As
reaction (\ref{HCN1}) proceeds, the HCN abundance increases.
CH$_3^+$ is strongly dependent on CH$_3$OH which is highly abundant
before 10$^4$ years and so remains in the gas phase with a  constant
column density of 10$^{12}$cm$^{-2}$ between 10$^2$ and 10$^4$
years. When the abundance of HCN increases, reaction
(\ref{CH3CNHreac}) proceeds and the abundance of CH$_3$CNH$^+$
increases, allowing the formation of CH$_3$CN via reaction
(\ref{CH3CNreac}).

\begin{figure}\label{figure4}
\includegraphics[width=85mm]{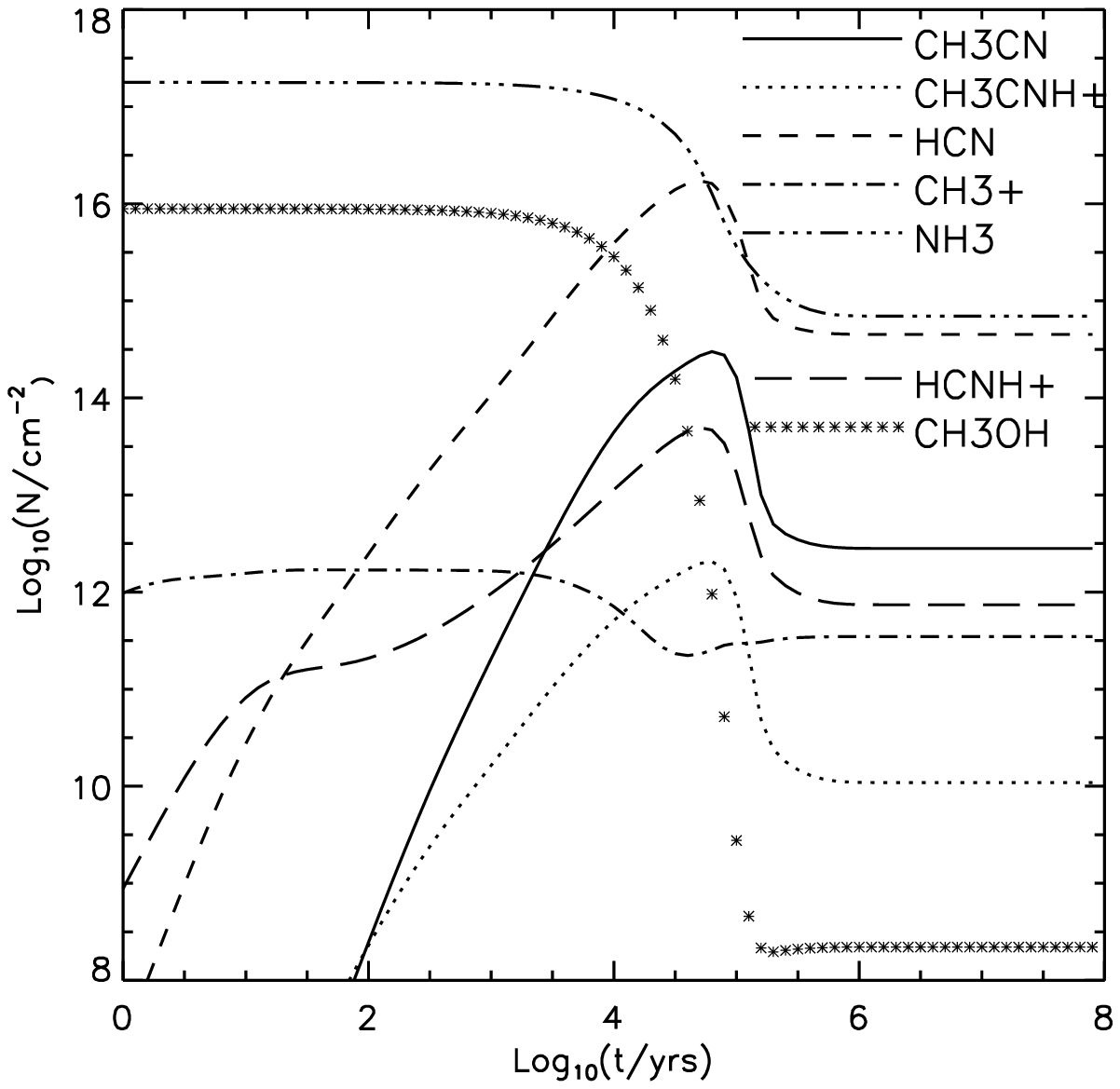}
 \caption{As for Figure  \ref{figure1}, but shows the parent species responsible for the formation of CH$_3$CN.}\label{figure4}
\end{figure}

\begin{figure}\label{figure5}
\includegraphics[width=85mm]{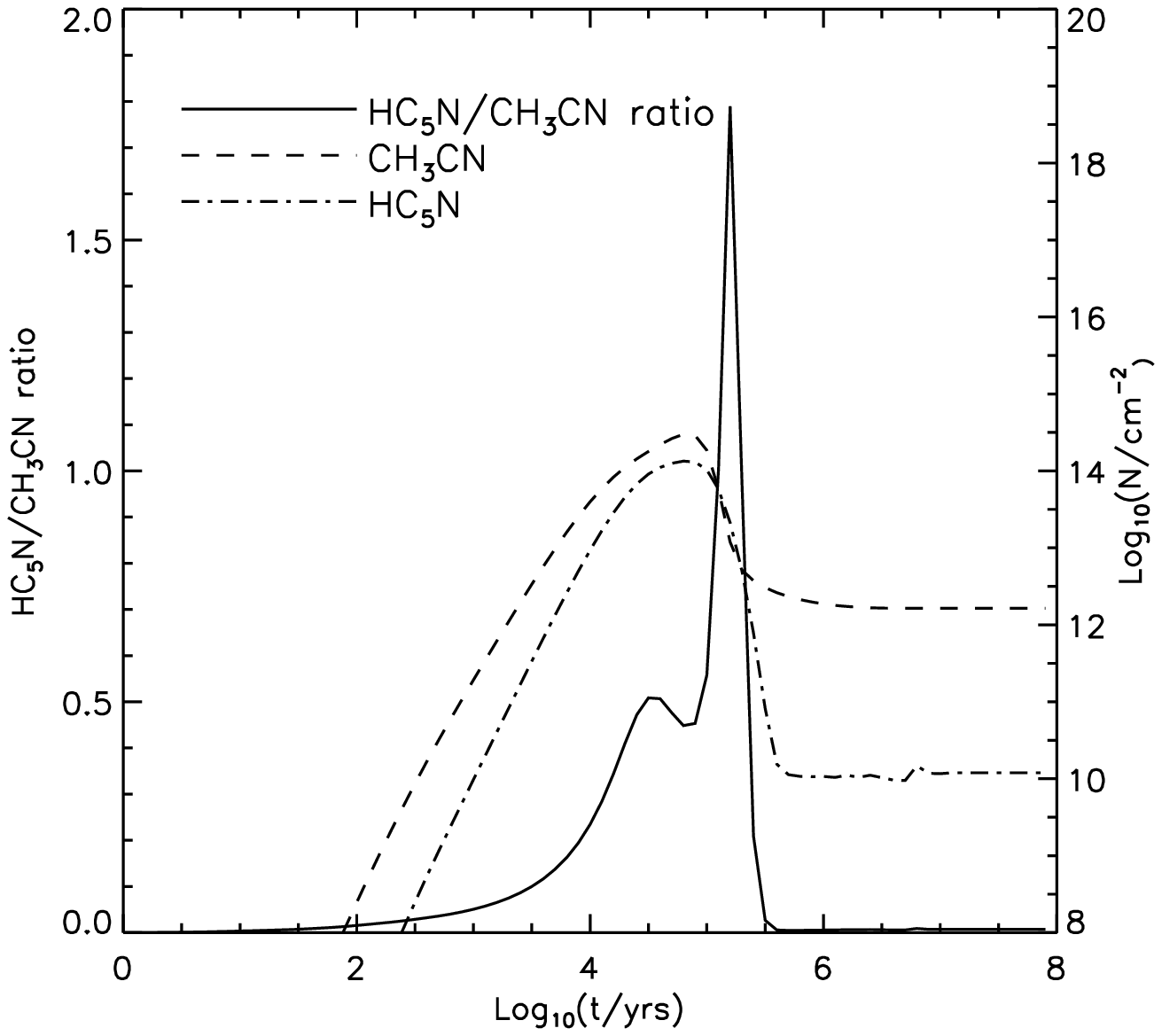}
 \caption{ Time dependent HC$_5$N/CH$_3$CN ratio for model in Figure 1.}\label{figure5}
\end{figure}

The abundances of HC$_5$N and CH$_3$CN are not directly linked via
their respective reaction networks, however the abundances of both
are strongly dependent on their parent grain mantle species. The
abundances of both increase at the onset of the rapid gas phase
chemistry caused by the destruction of the grain mantle species. As
a useful comparison, we plot the time dependent HC$_5$N/CH$_3$CN
ratio in Figure \ref{figure5}. The ratio increases considerably
between 10$^{4}$ and 10$^{5.5}$ years and so could prove useful in
determining the age of  hot cores within this range.










\section{Molecular line excitation model}\label{statequilsection}

\begin{figure*}
\includegraphics[width=140mm]{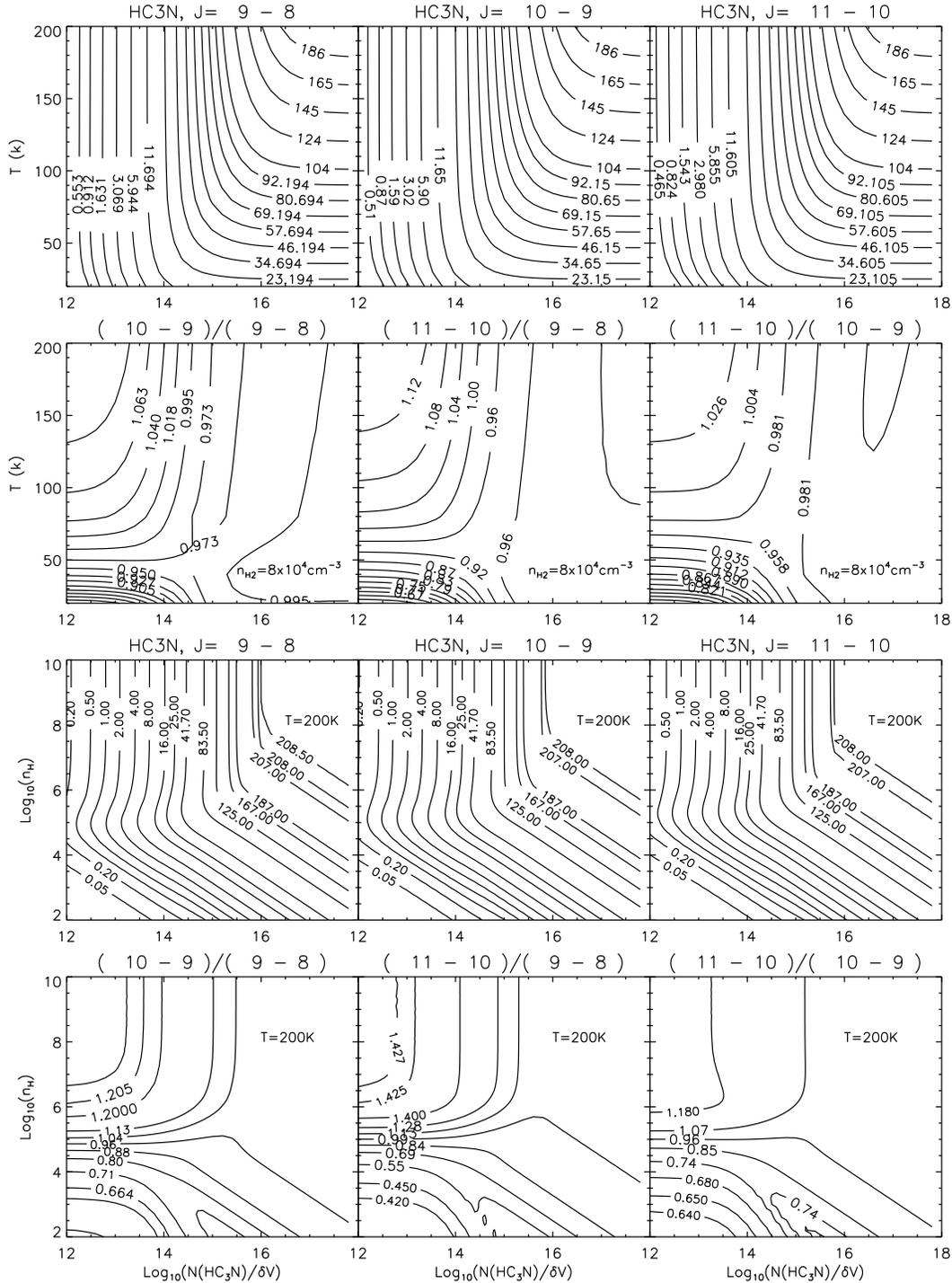}
\caption{Line intensities and ratios of HC$_3$N for the J=9-8
(81.88GHz), J=10-9 (90.98GHz)  and  J=11-10 (100.08GHz) lines
calculated by the excitation model. The first row shows the line
intensities in units of K km/s as a function of temperature and
column density (normalised by line width) assuming a density of
n$_{\mathrm{H2}}=8\times10 ^4$cm$^{-3}$. In the second row, three
line ratios as a function of temperature and column density for
constant density  8 $\times$$10^4$cm$^{-3}$ are plotted. In the
third and fourth rows, the corresponding plots as a function of
density for a constant temperature of 200K are
shown.}\label{HC3Nstat}
\end{figure*}

\begin{figure*}
\includegraphics[width=140mm]{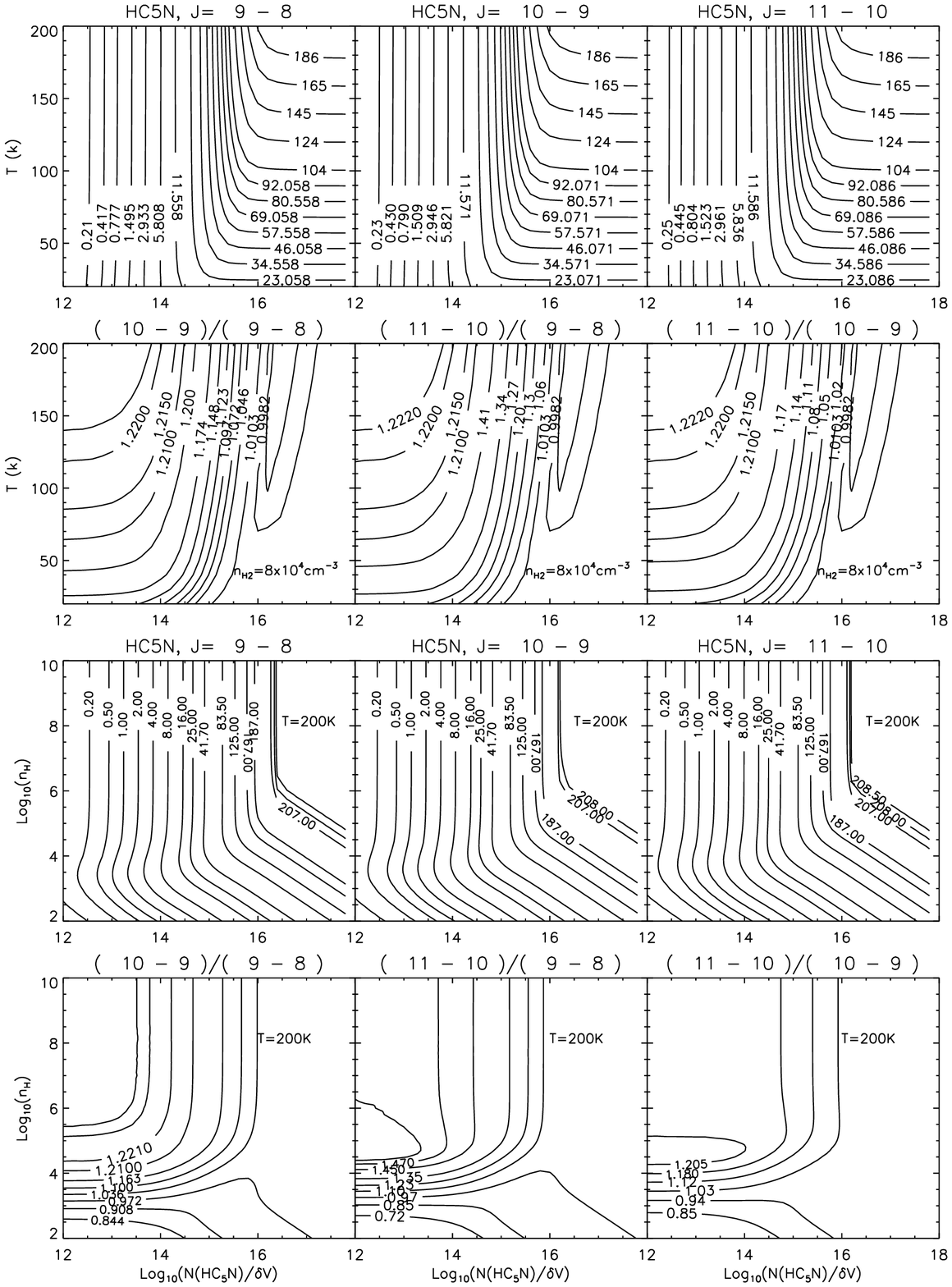}
\caption{As for Figure \ref{HC3Nstat} but for the J=9-8 (23.96GHz),
J=10-9 (26.63GHz) and J=11-10 (29.29GHz) lines of HC$_5$N.
}\label{HC5Nstat}
\end{figure*}







In this section, a molecular excitation model is used to calculate
the excitation of HC$_3$N,  HC$_5$N and HC$_7$N.  The collisional
rates (per second per molecule) with H$_2$  are taken from the LAMDA
database \citep{Setal05}. Only collision rates for the ground
vibrational state are currently available in the database and it
outside the scope of this work to consider the higher vibrational
transitions. We note however, that although higher vibrational lines
of HC$_3$N have been observed in SgB2 \citep{Goldetal83,
Goldetal85}, \cite{Walshetal07} did not detect any higher
vibrational transitions of HC$_3$N, HC$_5$N or HC$_7$N in G305A.
Here we only consider the ground vibrational states and will have to
wait until the collision rates are available for higher states to
update our model in the future.

The collision rates depend on the cross section of HC$_3$N and the
density of H$_2$.  The LAMDA database also contains the relevant
transitional frequencies, rotational levels and Einstein
A-coefficients for 24 molecules compiled from sources such as the
JPL Molecular Spectroscopy database \citep{Pickea98} and the CDMS
catalogue \citep{CDMS}.  The escape probability method is used here
to simplify the statistical equilibrium equations (see for example
\citealt{Setal05}), with the model outputting the intensity of a
line for a given density and temperature. Similar model calculations
for other species of interest can be found in, for example,
\cite{Liszt06}, \cite{Liqueetal06}, \cite{Danielcerndub06},
\cite{vandertakandhoger07}, and \cite{Walshetal07}.

In Figure \ref{HC3Nstat}, line intensities and line ratios are
plotted for the 81.88, 90.98 and  100.08GHz rotational lines of
HC$_3$N. The LAMDA database \citep{Setal05} contains collision rates
of the first 21 levels (9-181GHz) of HC$_3$N (however the higher
transitions are affected by truncation).  These plots are useful in
interpreting observations since a measured line intensity, or line
ratio can directly be compared on the plot, and the column density
and/or density may be determined. In the first row, the individual
line intensities in units K kms$^{-1}$ are given for a density of
n$_{\mathrm{H2}}=8\times10^4$cm$^{-3}$ as a function of temperature
and column density. Each plot shows the same overall trends, with
the intensities becoming relatively constant with temperature for
given normalised column densities  below 10$^{14}$cm$^{-2}$. A
single observed line intensity could be directly compared to these,
after telescope beam dilution effects have been taken into account.
As illustrated in, for example, \citep{Walshetal07}, plotting
multiple lines from non-LTE models allows for the determination of
the column density, as well as putting limits on the density and
temperature of the region.

We also show in the second row of Figure \ref{HC3Nstat} the line
ratios as a function of temperature for n$_\mathrm{H2}$ = 8
$\times$$10^4$cm$^{-3}$, which do not require any correction for
beam dilution effects. The intensity of the 100.08GHz (J=11-10) line
is strongest (for a given temperature and density), then the
90.98GHz (J=10-9) and 81.88GHz (J=9-8) lines. In the hot core
regime, for temperatures of the order 150K and above, the ratios are
relatively constant with column density and so potentially allowing
one to read off the column density directly using an observed line
ratio value.

 In the third row of Figure \ref{HC3Nstat}, the line
intensities are shown again, but this time as a function of density
for T=200K. Each contour becomes constant with column density as the
density increases, and the LTE approximation is approached with
increasing n$_\mathrm{H2}$.  One could safely assume the LTE
approximation, and use a rotation diagram, for densities above
10$^6$cm$^{-3}$ and if the line width normalised HC$_3$N column
density is expected to be in the range 10$^{12}$cm$^{-2}$  to
10$^{16}$cm$^{-2}$. Finally, in the bottom row of Figure
\ref{HC3Nstat}, the line ratios are plotted as a function of
density. As the column density decreases, the contours become
constant with density. The trend towards LTE with increasing density
is also clear.

Depending on what information is available, either series (row) of
plots in Figure \ref{HC3Nstat} may be the most useful in
interpreting observations.  If either the temperature or density is
known than the complexity is reduced greatly, and the corresponding
plots  with the appropriate temperature or density can be produced.
Notwithstanding, however, even if these are unknown, the combination
of all four plots can still be useful in determining limits on the
density, temperature and column density. Collision rates for the
larger cyanopolyynes  HC$_5$N, HC$_7$N, HC$_9$N etc. have yet to be
published. To calculate the excitation of HC$_5$N, we have used the
collisional rates of HC$_3$N \citep{Setal05}, however the cross
section for HC$_5$N is scaled by a factor of 1.5 to account for the
greater geometric length (e.g., \citealt{Averyetal79, Snelletal81}).
The energy levels and Einstein A coefficients for HC$_5$N have been
taken from the CDMS database \citep{CDMS}. The collision rates for
HC$_3$N are available for the first 21 lines and in the case of
HC$_5$N this includes all transitions from 2GHz to 54GHz. In Figure
\ref{HC5Nstat}  the corresponding line intensities and line ratios
are plotted  for the 23.96, 26.63 and 29.29GHz lines of HC$_5$N,
showing similar characteristics as for HC$_3$N. A similar
calculation has also been performed for HC$_7$N, after scaling the
collision rates of HC$_3$N by 2 to account for the larger cross
section of HC$_7$N (plots not shown here). These plots show similar
features as those for HC$_3$N and HC$_5$N.

\begin{figure}
\includegraphics[width=85mm]{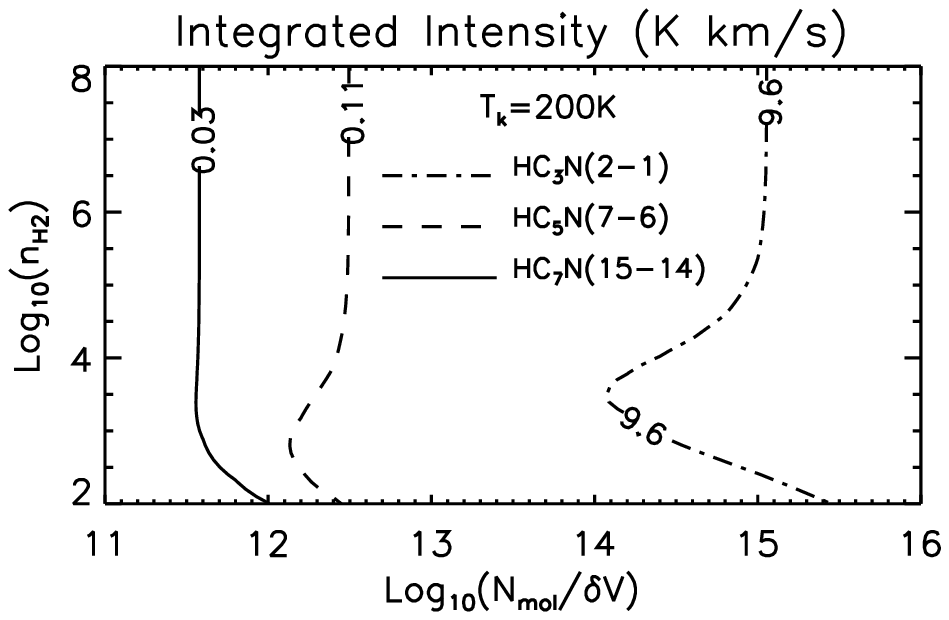}
\caption{Comparison of  cyanopolyynes from \citep{Walshetal07} using
a non-LTE excitation model. }\label{CyanObsfigure}
\end{figure}

In the ATCA observations of   \cite{Walshetal07} the  HC$_5$N(7-6)
18.64GHz and HC$_7$N(15-14) 16.92GHz lines  were not detected in
G305A, although Figure \ref{figure1} suggests that these species are
likely produced in this region.  We will now address this issue. The
1$\sigma$ upper limits of the integrated intensity of these
non-detections from the ATCA observations were $<$0.11Kkm/s and
$<$0.03Kkm/s, for HC$_5$N(7-6) and HC$_7$N(15-14), respectively.
Using the non-LTE model, we plot the upper limit intensities of each
of these lines in Figure \ref{CyanObsfigure} for a temperature of
200K, and varying density. We also plot the observed HC$_3$N(2-1)
18.20GHz line for comparison, this line was already considered in
the non-LTE analysis of \cite{Walshetal07}. For a density of
n$_\mathrm{H2}=8\times 10^4$cm$^{-3}$, the following limits are
imposed on the normalized column densities; N$_\mathrm{HC_5N}/\delta
V<  3.16 \times 10^{12}$ and N$_\mathrm{HC_7N}/\delta V< 3.89 \times
10^{11}$. In \cite{Walshetal07} an age range for  G305A of
$2\times10^4 < t_{core} <1.5 \times 10^5$ year was imposed, based on
all the detected lines. From Figure \ref{figure1}, this age range
implies column densities of
$10^{13}$cm$^{-2}<\mathrm{N}_{\mathrm{HC_5N}}<10^{14}$cm$^{-2}$ and
$2.5\times10^{12}$cm$^{-2}<\mathrm{N}_{\mathrm{HC_7N}}<6.3\times10^{13}$cm$^{-2}$,
with the lower limits defined by the later core age of $1.5\times
10^5$ years. After, assuming a line width of $>3.2$km/s for the
lines in Figure \ref{CyanObsfigure}, which is not unreasonable, the
column densities of HC$_5$N and HC$_7$N agree with Figure
\ref{figure1} for a core range of the order $1.5\times 10^5$ years.
The NH$_3$ observations of \cite{Walshetal07}  favor a later age for
G305A of the order 10$^5$ years, also. Around this time the
cyanopolyynes are being destroyed, and their column densities
decrease rapidly. This could explain why these species were not
detected in \cite{Walshetal07}, since there abundance was not high
enough to be detected.

\section{Discussion}\label{discussion}

The results from the chemical model in Sections
\ref{chemicalmodelsection} and \ref{chemicalmodelsection-cyan} show
that the  cyanopolyynes  switch on after 10$^2$ years and peak
around 10$^{4.5}$ years, before being rapidly destroyed before
10$^{5.5}$ years. Acetylene was shown to be a key species in the
formation of the cumulene carbenes H$_2$C$_n$, which react with CN
to form the cyanopolyynes. Since acetylene is abundant in large
quantities, as it has been evaporated from grain mantles, and CN is
formed easily in the gas phase, the cyanopolyynes should also be
found in hot cores. The formation and subsequent destruction of the
longer chain cyanopolyynes are relatively short lived, which may
explain why these are not often observed in  hot core sources.
However, if they are observed, they may prove very useful as
chemical clocks, as they could help pinpoint the age of the hot
core. Their detection in sources with similar physical conditions to
those of G305A in Section \ref{chemicalmodelsection} should imply
ages of the order of 10$^{4.5}$ years.

\begin{figure}
\includegraphics[width=85mm]{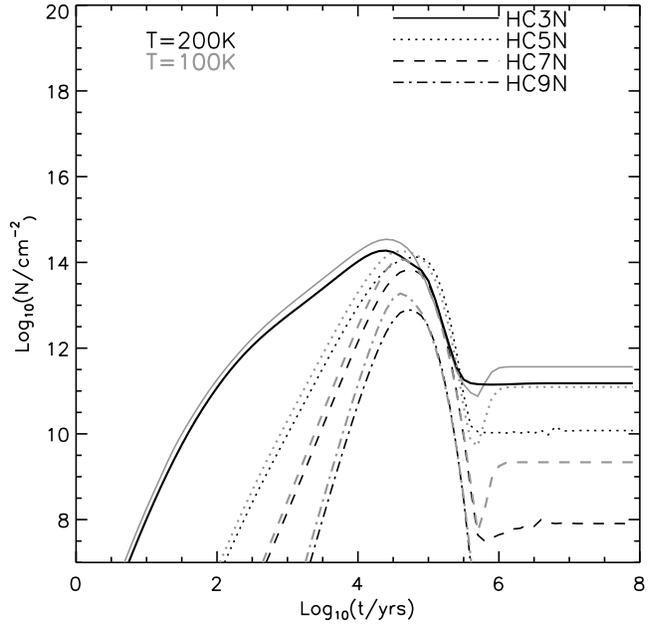}
\caption{Comparing the time dependent chemistry for model with
n$_\mathrm{H2}$ = 8 $\times$$10^4$cm$^{-3}$ and temperatures of 100K
and  200K. }\label{comparetemp}
\end{figure}

The calculated time dependent chemistry in Section
\ref{chemicalmodelsection} are for specific physical conditions;
n$_{\mathrm{H2}}=8\times 10^4$cm$^{-3}$ and T=200K. We now examine
the sensitivity of the chemical model results to changes in density
and temperature so that these results may be extended to hot cores
with different physical conditions to G305A.  Since the conditions
inside the core are constant throughout, and the source fills the
beam and beam dilution effects are not so important, changes in the
size of the core,  will only scale the column densities and not
change the time dependent features.  Since the reaction rates are
temperature dependent, a change in temperature will effect the model
results. The reaction rates for two body reactions are of the form
\begin{equation}
k=\alpha(T/300)^\beta \mathrm{exp}(\gamma/T) \,\,\,\,
\mathrm{cm^3s^{-1}}
\end{equation}
where $\alpha$, $\beta$ and $\gamma$ are constants taken from the
UMIST database \citep{Wetal05}. The reaction rates for reaction
(\ref{cyanreac}), are the same regardless of the length of each
cyanopolyyne but vary with temperature. A reduction in temperature
to 100K, increases the reaction rate by 1.6 and therefore increases
the abundances. This is shown in Figure \ref{comparetemp}.

\begin{figure}
\includegraphics[width=85mm]{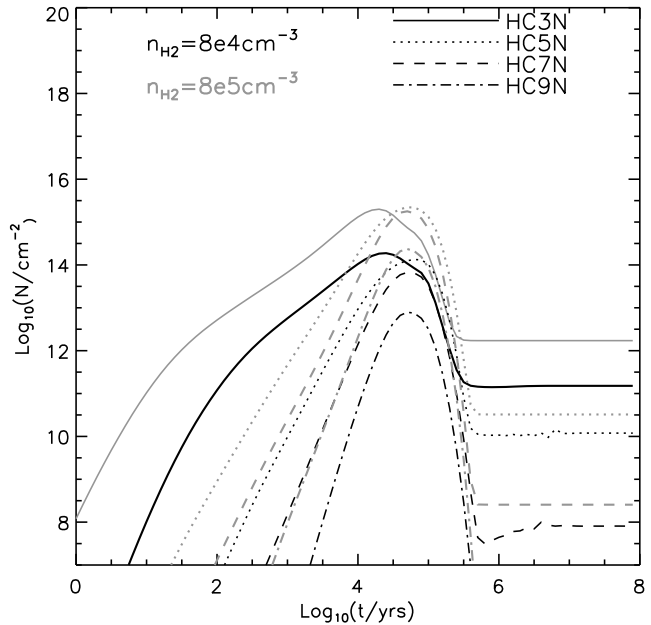}
\caption{Comparing the time dependent chemistry for model with a
temperature of 200K and densities of  n$_\mathrm{H2}$ = 8
$\times$$10^4$cm$^{-3}$ and n$_\mathrm{H2}$ = 8
$\times$$10^5$cm$^{-3}$. }\label{comparedens}
\end{figure}

Since we have only considered the physical conditions for
G305.2+0.2, we now increase the density to illustrate that longer
chain cyanopolyynes should also be formed in higher density cores.
In Figure \ref{comparedens} we consider an increase the  density by
a factor of 10 to n$_{\mathrm{H2}}=8\times10^5$cm$^{-3}$. The
species abundances are enhanced by at least a factor of 10. The time
dependent features follow the same movements as for the lower
density G305A case.

\section{Summary}\label{summary}

In the core chemical model presented here, with density
n$_\mathrm{H2}=8\times10^4$cm$^{-3}$ and a temperature of 200K, the
cyanopolyynes show highly time dependent chemistry between 10$^2$
and 10$^6$ years. Although the larger cyanopolyynes HC$_\mathrm{n}$N
(n$>3$) are usually not observed and have not been considered in
theoretical models in the past, we have shown that they can be
produced under hot core conditions. Their abundance is directly
linked to acetylene, which is highly abundant in the earlier stages
of the core evolution when it is evaporated from grain mantles.  The
reactions responsible for processing acetylene into the daughter
species cyanopolyynes are efficient under the physical conditions of
G305A, as well as the higher density and temperature cases
considered in Section \ref{discussion}. The abundance of the larger
cyanopolyynes increase and decrease over a short period of the order
of 10$^{2.5}$ years, and are relatively short lived compared with
other more commonly observed species. This may explain why they have
been observed in so few sources.  Since they are only present in the
gas phase for a relatively short time, they make useful species as
chemical clocks, if the source age lies within this window. HC$_3$N,
however, is produced in large abundances much earlier on than the
larger cyanopolyynes which may explain why it has been detected in
several sources.

Through the use of a molecular excitation model, we presented a
series of line intensity and line ratio plots, useful for the
interpretation of spectral line data of HC$_3$N and HC$_5$N. Given a
series of line intensities measured from a source, these and  the
associated line ratios can be used to look up the column density of
a species.  Although there were non-detections of HC$_5$N and
HC$_7$N in \cite{Walshetal07}, we show that this is most probably
due to these species having low column densities which are below the
observing range of the ACTA.

\section*{Acknowledgments}
We would like to thank the anonymous referee who has helped improve
the quality of this paper. This research was supported by the
Australian Research Council. Astrophysics at QUB is supported by a
grant from the STFC.


 \label{lastpage}

\end{document}